\documentstyle[12pt]{article}


\newlength{\minitwocolumn}
\catcode`\@=11
\long\def\@makefntext#1{
\protect\noindent \hbox to 3.2pt {\hskip-.9pt  
$^{{\eightrm\@thefnmark}}$\hfil}#1\hfill}               

\def\@makefnmark{\hbox to 0pt{$^{\@thefnmark}$\hss}}    
        
\def\ps@myheadings{\let\@mkboth\@gobbletwo
\def\@oddhead{\hbox{}
\rightmark\hfil\eightrm\thepage}   
\def\@oddfoot{}\def\@evenhead{\eightrm\thepage\hfil
\leftmark\hbox{}}\def\@evenfoot{}
\def\sectionmark##1{}\def\subsectionmark##1{}}

\font\eightrm=cmr8

\font\sc=cmr5 scaled\magstep1

\def\MPL{Mod.~Phys.~Lett. }

\def\PL{Phys.~Lett. }
\def\PR{Phys.~Rev. }

\def\PTP{Prog.~Theor.~Phys. }

\def\delzero{\delta_0}
\def\brs{\delta}

\newcommand\gh{{\rm gh}}
\newcommand\antigh{{\rm antigh}}

\newcommand{\rd}{\overleftarrow{\partial}} 
\newcommand{\ld}{\overrightarrow{\partial}} 
\newcommand{\abv}[2]{{({\,{#1}\,,\,{#2}\,})_{\hbox{\sc AB}}}}
\newcommand{\sbv}[2]{{({\,{#1}\,,\,{#2}\,})}}
\newcommand{\bracket}[2]{\langle #1\,,#2\rangle}

\def\bF{\mbox{\boldmath $F$}}
\def\bG{\mbox{\boldmath $G$}}

\def\bb{\mbox{\boldmath $B$}}
\def\bPhi{\mbox{\boldmath $\Phi$}}
\def\bh{\mbox{\boldmath $h$}}

\newcommand{\aone}{{\tilde A_1}{}}
\newcommand{\azero}{{\tilde \alpha_0}{}}
\newcommand{\btwo}{{\tilde B_2}{}}
\newcommand{\beone}{{\tilde \beta_1}{}}
\newcommand{\bone}{{\tilde B_1}{}}
\newcommand{\bezero}{{\tilde \beta_0}{}}
\newcommand{\phizero}{{\tilde \phi}{}}
\newcommand{\dtwo}{{d}{}}

\begin{document}


\baselineskip 0.7cm

\begin{titlepage}
\begin{flushright}
UT-04-20
\end{flushright}

\vskip 1.35cm
\begin{center}
{\Large \bf
Dimensional Reduction of Nonlinear Gauge Theories}
\vskip 1.2cm
Noriaki IKEDA$^1$%
\footnote{E-mail address:\ ikeda@yukawa.kyoto-u.ac.jp}
\ and IZAWA K.-I.$^2$
\vskip 0.4cm
{\it $^1$Department of Mathematical Sciences, 
Ritsumeikan University \\
Kusatsu, Shiga 525-8577, Japan }\\
{\it $^2$Department of Physics,
University of Tokyo \\
Tokyo 113-0033, Japan }\\

\date{}

\vskip 1.5cm

\begin{abstract}
We investigate an extension of 2D nonlinear gauge theory
from the Poisson sigma model based on Lie algebroid
to a model with additional two-form gauge fields.
Dimensional reduction of 3D nonlinear gauge theory
yields an example of such a model, which provides
a realization of Courant algebroid by
2D nonlinear gauge theory.
We see that the reduction of the base structure
generically results in a modification of
the target (algebroid) structure.
\end{abstract}
\end{center}
\end{titlepage}

\setcounter{page}{2}


\rm
\section{Introduction}
\noindent
Topological gauge field theories of the Schwarz type
\cite{Bir}
are grouped into two categories:
Chern-Simons-BF gauge theory
\cite{Bir}
and nonlinear gauge theory
\cite{Ike,Iza}.
Topological nonlinear gauge theory
can be obtained as deformations of the former
\cite{Izaw}
and may be regarded as
the most general form of the topological gauge field theories
of the Schwarz type.

Two-dimensional nonlinear gauge theory of the simplest form
\cite{Iza}
turns out to be determined solely by the Lie algebroid structure
\cite{Lev},
the structure functions of nonlinear Lie (finite W
\cite{Boe})
algebra, or the data of Poisson algebra
\cite{Str},
which is manifested
in the name of the Poisson sigma model
\cite{Sch}.%
\footnote{Since it has scalar fields in its field contents,
it is apparently a sigma model as well as a gauge theory.}

Various extensions of the 2D nonlinear gauge theory may be considered.
In fact, 2D theory based on graded structures was obtained
\cite{Iked}
soon after the exposition of the original bosonic model.
Higher-dimensional generalization of the 2D nonlinear gauge theory
is also possible
\cite{Izaw}.
Accordingly, 3D nonlinear gauge theory is systematically constructed
\cite{Ikeda:2000yq,Ikeda:2001fq,Ikeda:2002wh,Ikeda:2002qx}%
\footnote{Yet higher-dimensional cases are also considered
in Ref.\cite{Izaw,Ikeda:2001fq,Ikeda:2002qx}.}\,\,
through deformations of Chern-Simons-BF gauge theories
and the Courant algebroid
\cite{Courant}
structure underlying it is identified
\cite{Ikeda:2002wh,Ikeda:2002qx}.

In this paper,
we investigate an extension of 2D nonlinear gauge theory
from the Poisson sigma model based on the Lie algebroid
to a model with additional two-form gauge fields.
This extension is motivated by considering
dimensional reduction of 3D nonlinear gauge theory
based on the Courant algebroid.

\section{2D Nonlinear Gauge Theory with Two-Forms}
\noindent
Let us first present an extended action of 2D nonlinear gauge theory
postponing its derivation to the following sections.

We consider an action with $2$-forms $\btwo_A$,
\begin{eqnarray}
S = \int_{\Sigma} h_A \dtwo \Phi^A + \frac{1}{2} W^{AB} (\Phi) h_A h_B
+ V^A (\Phi) \btwo_A,
\label{nlaction}
\end{eqnarray}
where $\Sigma$ denotes a two-dimensional base manifold and 
$M$ a target space manifold of
a smooth map $\Phi:\Sigma \to M$ with a local
coordinate expression $\{\Phi^A\}$.
Here $h_A$ is a section of $T^*\Sigma \otimes \Phi^*(T M)$,
$\btwo_A$ is a section of $ \wedge^2 T^*\Sigma \otimes \Phi^*(T M)$,
and $W^{AB}=-W^{BA}$ denotes
a bivector field and $V^A$ a vector field on the target space $M$.
The last term with 2-forms as Lagrange multiplier fields
was previously considered by Batalin and Marnelius
\cite{Mar}.
When $V^A(\Phi) \equiv 0$, the action (\ref{nlaction}) reduces to the
Poisson sigma model.

The action has the following gauge symmetry:
\begin{eqnarray}
&& \brs \Phi^A = - W^{AB} c_B,
\nonumber \\
&& \brs h_A = \dtwo c_A 
+ \frac{\partial W^{BC}}{\partial \Phi^A} h_B c_C
- \frac{\partial V^B}{\partial \Phi^A} t_B,
\nonumber \\
&& \brs \btwo_A = \dtwo t_A + U_A{}^{BC} (h_B t_C 
- \btwo_C c_B) + \frac{1}{2} X_A{}^{BCD} h_B h_C c_D,
\label{gaugebrs}
\end{eqnarray}
if 
$U_A{}^{BC}(\Phi)$ and $X_A{}^{BCD}(\Phi)$
satisfy the identities
\begin{eqnarray}
&& W^{D[A} \frac{\partial W^{BC]}}{\partial \Phi^D} = 
V^D X_D{}^{ABC},
\label{jacobi}
\nonumber \\
&& W^{AB} \frac{\partial V^C}{\partial \Phi^A} + U_A{}^{BC} V^A = 0,
  \label{defu}
\end{eqnarray}
where $c_A$ is a $0$-form gauge parameter and $t_A$ is a $1$-form
gauge parameter
with $X_A{}^{BCD}$ completely antisymmetric with respect to
the indices $BCD$.

In fact, the action $S$ is gauge invariant 
\begin{eqnarray}
\brs S = \int_{\Sigma} \dtwo( c_A \dtwo \Phi_A + V^A t_A),
\end{eqnarray}
and its equations of motion are given by
\begin{eqnarray}
&& \dtwo h_A + \frac{1}{2} \frac{\partial W^{BC}}{\partial \Phi^A} h_B
h_C + \frac{\partial V^B}{\partial \Phi^A} {\tilde B}_{2B} =0, 
\nonumber \\
&& \dtwo \Phi^A + W^{AB} h_B = 0,
\nonumber \\
&& V^A =0.
\label{eom}
\end{eqnarray}

If we can take $V^A=(0, V^a(\Phi^A))$ for $\{a\} \subset \{A\}$
with $V^a(\Phi^A)={\Phi}^a$
through a coordinate transformation on $M$,%
\footnote{This is possible locally for generic $V^a(\Phi^A)$.}\,\, 
we may locally eliminate
the fields ${\Phi}^a$ by means of the equations of motion
(let us call the extended theory {\it reducible} in this case).
Then the theory reduces to the Poisson sigma model with
the target space dimension reduced accordingly.

\section{Derivation of the 2D Theory with Two-Forms}
\noindent
In this section, we derive the action given in the previous
section through a deformation of BF theory in two dimensions.
We first set up a superformalism
\cite{Bat,Cat}
of the BF theory
and then perform a consistent deformation
\cite{BH}
thereof.

\subsection{Superformalism of 2D BF theory with two-forms}
\noindent
We start with a free action
\begin{eqnarray}
S_{\hbox{\sc A}}
= \int_{\Sigma} 
h_A \dtwo \Phi^A,
\label{abelianaction}
\end{eqnarray}
where 
$h_A$ is a $1$-form gauge field and
$\Phi^A$ is a $0$-form scalar field on $\Sigma$.
We further introduce $\btwo_A$ as a $2$-form field on $\Sigma$.
This action has an abelian gauge symmetry
\begin{eqnarray}
&& \delzero h_A = d c_A, \nonumber \\
&& \delzero \btwo_A = d t_A,
\label{abrs}
\end{eqnarray}
where $c_A$ is a $0$-form gauge parameter and $t_A$ is a $1$-form
gauge parameter. 
The gauge symmetry for $\btwo_A$ is trivially satisfied
and reducible.
We need the following tower of the `ghosts for ghosts' to analyze the
complete gauge degrees of freedom:
\begin{eqnarray}
&& \delzero h_A = d c_A,
\qquad 
\delzero c_A = 0,
 \nonumber \\
&& \delzero \btwo_A = d t_A,
\qquad \delzero t_A = d v_A,
\qquad \delzero v_A = 0,
\label{agauge}
\end{eqnarray}
where $v_A$ is a $0$-form gauge parameter.

Let us here set up
the Batalin-Vilkovisky formalism
\cite{Bat}
to adopt the Barnich-Henneaux approach for consistent deformation
\cite{BH}
in the next subsection.
First we take $c_A$ to be the
FP ghost $0$-form with the ghost number $1$, $t_A$ 
to be a $1$-form with the ghost number $1$, and 
$v_A$ to be a $0$-form with the ghost number $2$.
Next we introduce the antifields for all the fields:
$\varphi^+$ denotes the antifield for the field $\varphi$.
Note that such relations as
${\rm gh}(\varphi) + {\rm gh}(\varphi^+) = -1$ and
${\rm deg}(\varphi) + {\rm deg}(\varphi^+) = 2$ (in two dimensions)
are imposed,
where ${\rm gh}(\varphi)$ and 
${\rm gh}(\varphi^+)$ are the ghost numbers
of the (anti)fields $\varphi$ and $\varphi^+$ and
${\rm deg}(\varphi)$ and ${\rm deg}(\varphi^+)$ 
are their form degrees,
respectively.

In order to simplify
combinatorics, we adopt a superformalism
\cite{Cat}.
Namely, we combine the fields, antifields,
and their gauge descendant fields
as superfield components:
\begin{eqnarray}
\bh_A &=& c_A + h_A + \Phi^+_A,
\nonumber \\
\bPhi^A & = & \Phi^A + h^{+A} + c^{+A},
\nonumber \\
\bb_A &=& v_A + t_A + \btwo_A,
\nonumber \\
\bb^{+A} & = & \btwo^{+A} + t^{+A} + v^{+A}.
\label{component}
\end{eqnarray}
Note that the component fields $F$ in a superfield have the common
{\it total degree} $|F| \equiv \gh F + \deg F$.
The total degrees of $\bh_A$, $\bPhi^A$, $\bb_A$, and $\bb^{+A}$ 
are $1$, $0$, $2$, and $-1$, respectively.
We introduce the {\it antighost number} $\antigh (\bF)$ of 
a superfield $\bF$ with only $\antigh (\bb^{+A}) = 1$ nonvanishing.
The {super antibracket} of two superfields
$\bF$ and $\bG$ is given by
\begin{eqnarray}
\sbv{\bF}{\bG} &=& 
\bF \cdot \frac{\rd}{\partial \bPhi^A} \cdot
\frac{\ld }{\partial \bh_A} \cdot \bG
- 
\bF \cdot \frac{\rd}{\partial \bh_A} \cdot
\frac{\ld }{\partial \bPhi^A} \cdot \bG
\nonumber \\
&& 
+ \bF \cdot \frac{\rd }{\partial \bb_A} \cdot
\frac{\ld }{\partial \bb^{+A}} \cdot \bG
- \bF \cdot \frac{\rd}{\partial \bb^{+A}} \cdot 
\frac{\ld }{\partial \bb_A} \cdot \bG,
\label{bvbracket}
\end{eqnarray}
where we have utilized the {\it super product},
the {\it super antibracket}, and the {\it super differentiation} defined
in the Appendix.

We can now construct the Batalin-Vilkovisky action to the original action 
(\ref{abelianaction}) with the superfields as follows:
\begin{eqnarray}
S_0 &=& \int_{\Sigma} 
\bh_A \cdot \dtwo \bPhi^A + \bb^{+A} \cdot \dtwo \bb_A,
\nonumber \\
&=& \int_{\Sigma} 
h_A \dtwo \Phi^A - c_A \dtwo h^{+A} 
- \btwo^{+A} \dtwo t_A + t^{+A} \dtwo v_A,
\label{abelianbvaction}
\end{eqnarray}
where only the $2$-form part of the integrand survives integration
on the two-dimensional manifold $\Sigma$.
The total degree of the integrand of $S_0$ is two and its antighost
number is zero.
If we set all the antifields equal to zero,
Eq.(\ref{abelianbvaction}) reduces to Eq.(\ref{abelianaction}).

If we assigned
the total degrees of superfields
$\bh_A$, $\bPhi^A$, $\bb_A$, and $\bb^{+A}$ 
as $1$, $0$, $0$, and $1$, respectively, 
we could regard the superfield action (\ref{abelianbvaction}) as 
a Batalin-Vilkovisky action to the usual abelian BF theory.
In fact, there is a one-to-one map
in terms of superfield actions
from the 2D abelian BF theory with two-forms
to the 2D abelian BF theory,
which changes the total degrees of superfields $\bb_A$ and $\bb^{+A}$ 
from $2$ and $-1$ to $0$ and $1$, respectively.

The BRST transformation of the superfield $\bF$ for the
above action is given by
\begin{eqnarray}
\delzero \bF \equiv \sbv{S_0}{\bF},
\end{eqnarray}
which yields
\begin{eqnarray}
&& \delzero \bh_A
= \sbv{S_0}{\bh_A} = \dtwo \bh_A,
\nonumber \\
&& \delzero \bPhi^A
= \sbv{S_0}{\bPhi^A} = \dtwo \bPhi^A,
\nonumber \\
&& \delzero \bb_A
= \sbv{S_0}{\bb_A} = \dtwo \bb_A,
\nonumber \\
&& \delzero \bb^{+A}
= \sbv{S_0}{\bb^{+A}} = \dtwo \bb^{+A}.
\label{ababelBRST}
\end{eqnarray}
By expanding these BRST transformations 
to the components Eq.(\ref{component}), 
we obtain the BRST transformation of each field and antifield
as follows:
\begin{eqnarray}
&& 
\delzero \Phi_A^{+} = \dtwo h_A,
\qquad
\delzero h_A = \dtwo c_A,
\qquad 
\delzero c_A = 0,
 \nonumber \\
&& 
\delzero c^{+A} = \dtwo h^{+A},
\qquad
\delzero h^{+A} = \dtwo \Phi^A,
\qquad 
\delzero \Phi^A = 0,
 \nonumber \\
&& 
\delzero \btwo_A = \dtwo t_A,
\qquad 
\delzero t_A = \dtwo v_A,
\qquad 
\delzero v_A = 0,
 \nonumber \\
&& 
\delzero v^{+A} = \dtwo t^{+A},
\qquad 
\delzero t^{+A} = \dtwo \btwo^{+A},
\qquad 
\delzero \btwo^{+A} = 0,
\label{gauge}
\end{eqnarray}
which reproduces the original gauge transformation (\ref{agauge}).
It is simple to confirm that $S_0$ is BRST invariant and 
$\delzero^2=0$ on all the fields.

\subsection{Consistent deformation of the Batalin-Vilkovisky action}
\noindent
Let us consider a deformation of the action $S_0$,
\begin{eqnarray}
&& S = S_0 + g S_1 + g^2 S_2 + \cdots,
\label{pertur}
\end{eqnarray}
where $g$ is a deformation parameter or a coupling constant of the
theory. 

For a consistent deformation
\cite{BH},
we demand the total action $S$ to satisfy the classical master
equation
\begin{eqnarray}
\sbv{S}{S} = 0.
\label{master}
\end{eqnarray}
Substituting Eq.(\ref{pertur}) to Eq.(\ref{master}), we obtain the $g$ power 
expansion of the master equation
\begin{eqnarray}
\sbv{S}{S} = 
\sbv{S_0}{S_0} 
+ 2g \sbv{S_0}{S_1}
+ g^2 [ \sbv{S_1}{S_1} + 2 \sbv{S_0}{S_2} ] + {\cal O}(g^3) = 0.
\label{purmaster}
\end{eqnarray}

We solve this equation order by order
with physical requirements for the solutions:
We require the Lorentz invariance (Lorentzian case)
or $SO(2)$ invariance (Euclidean case)
of the action.
We assume that $S$ is {\it local},
that is, $S$ is given by the integration of a {local}
Lagrangian:
$
S = \int_{\Sigma} {\cal L}.
$
Note that we exclude the solutions whose BRST transformations
are not deformed ($\brs=\delzero$) as trivial ones.
This condition is implied by the assumption that each term contains
at least one antifield in $S_i$ for $i \geq 1$.

At the $0$-th order, we obtain $\delzero S_0 = \sbv{S_0}{S_0} =0$, which
is already satisfied.
At the first order of $g$ in Eq.~(\ref{purmaster}),
\begin{eqnarray}
\delzero S_1 = \sbv{S_0}{S_1} =0
\label{1brst}
\end{eqnarray}
is required. 
$S_1$ is given by the integration of a {local}
Lagrangian from the assumption:
\begin{eqnarray}
&&
S_1 = \int_{\Sigma} {\cal L}_1,
\label{s1act} 
\end{eqnarray}
where ${\cal L}_1$ can be constructed from the superfields 
$\bh_A$, $\bPhi^A$, $\bb_A$, and $\bb^{+A}$.
If a monomial in ${\cal L}_1$ includes a differential $\dtwo$, 
it is proportional to the free equations of motion.
Therefore it can be absorbed into the abelian action
(\ref{abelianbvaction}) through local field redefinitions of
superfields and these terms are BRST trivial in the BRST cohomology.
Hence the nontrivial deformation does not include
the differential $\dtwo$, and thus
${\cal L}_1$ is a degree two function of the superfields
$\bh_A$, $\bPhi^A$, $\bb_A$, and $\bb^{+A}$.

At the second order of $g$, 
\begin{eqnarray}
\sbv{S_1}{S_1} + 2 \sbv{S_0}{S_2} = 0
\label{pursec}
\end{eqnarray}
is required.
We cannot construct nontrivial $S_2$ to satisfy Eq.(\ref{pursec})
from the integration of a local Lagrangian, since
$\delzero$-BRST transforms of the local terms are always total
derivative.
Therefore, if we assume locality of the action, $S_2$ is BRST
trivial (the Poincar\'e lemma),
that is, we obtain the relation $\sbv{S_0}{S_2} = 0$ and 
we can absorb $S_2$ into $S_1$.
Similarly, when we solve the higher order $g$ part of
Eq.(\ref{purmaster}) recursively,
we find that $S_i$ is BRST trivial for $i \geq 2$.
Hence we may set $S_i = 0$ for $i \geq 2$. 
Then the condition (\ref{master}) reduces to
\begin{eqnarray}
\sbv{S_1}{S_1} = 0.
\label{s1s1}
\end{eqnarray}
This equation imposes conditions on the structure functions 
$f_i(\bPhi)$ in Eq.(\ref{s1k}).

Let us solve the condition (\ref{s1s1}) explicitly.
We expand $S_1$ by the antighost number 
\begin{eqnarray}
&&
S_1 = \sum_k S_1^{(k)} = \sum_k \int_{\Sigma} {\cal L}_1^{(k)},
\label{s1actk} 
\end{eqnarray}
where $S_1^{(k)}$ is the antighost number $k$ part of the action $S_1$ and 
${\cal L}_1^{(k)}$ is its Lagrangian ($k \geq 0$).
We can write the candidate ${\cal L}_1^{(k)}$ 
under the requirement $|{\cal L}_1^{(k)}|=2$ as
\begin{eqnarray}
&& 
{\cal L}_1^{(0)}
= \frac{1}{2} f_1^{AB} (\bPhi) \cdot \bh_A \cdot \bh_B
+ f_2^{A}(\bPhi) \cdot \bb_A, 
\nonumber \\
&& 
{\cal L}_1^{(1)}
= \frac{1}{3!} f_{3A}{}^{BCD} (\bPhi) \cdot \bb^{+A} \cdot \bh_B 
\cdot \bh_C \cdot \bh_D
+ f_{4A}{}^{BC} (\bPhi) \cdot \bb^{+A} \cdot \bh_B \cdot \bb_C, 
\label{s1k}
\end{eqnarray}
and so on, where $f_i (\bPhi)$ is a function of $\bPhi^A$.

We also expand $\sbv{S_1}{S_1}$ by the antighost number as follows:
\begin{eqnarray}
\sbv{S_1}{S_1} = \sum_{k} {\sbv{S_1}{S_1}}^{(k)} = 0,
\label{s1s1k}
\end{eqnarray}
where ${\sbv{S_1}{S_1}}^{(k)}$ is the antighost number $k$ part of 
$\sbv{S_1}{S_1}$.
This equation requires ${\sbv{S_1}{S_1}}^{(k)} = 0$
for all the nonnegative integers $k$,
and we can determine the conditions among $f_i$ recursively:
First, we consider $k=0$ part. 
We substitute Eq.(\ref{s1k}) to ${\sbv{S_1}{S_1}}^{(0)} = 0$ and 
obtain
\begin{eqnarray}
&& f_1^{D[A} \frac{\partial f_1^{BC]}}{\partial \Phi^D} 
+ f_2^D f_{3D}{}^{ABC} = 0,
\nonumber \\
&& - f_1^{AB} \frac{\partial f_2^C}{\partial \Phi^A} 
+ f_{4A}{}^{BC} f_2^A = 0.
  \label{fidentity}
\end{eqnarray}
As for the higher order terms with respect to the antighost number, 
we obtain conditions of higher order $f_i$'s.
It is sufficient to consider $k \leq 1$ terms
in order to obtain the deformed action and deformed BRST
transformation,
since higher order terms vanish when we set all the antifields equal to zero.

The BRST transformation of each superfield is given by
\begin{eqnarray}
&& \brs \bh_A = \dtwo \bh_A 
+ \frac{1}{2} g \frac{\partial f_1^{BC}}{\partial \bPhi^A} \cdot \bh_B
\cdot \bh_C
+ g \frac{\partial f_2^{B}}{\partial \bPhi^A} \cdot \bb_B
+ \frac{1}{3!} g \frac{\partial f_{3B}{}^{CDE}}{\partial \bPhi^A}
\cdot \bb^{+B} \cdot \bh_C \cdot \bh_D \cdot \bh_E
\nonumber \\
&& \qquad \quad
+ \frac{1}{2} g \frac{\partial f_{4B}{}^{CD}}{\partial \bPhi^A}
\cdot \bb^{+B} \cdot \bh_C \cdot \bb_D + \cdots,
\nonumber \\
&& \brs \bPhi^A = \dtwo \bPhi^A - gf_1^{AB} \cdot \bh_B 
+ \frac{1}{2} gf_{3B}{}^{ACD} \cdot \bb^{+B} \cdot \bh_C \cdot \bh_D
- gf_{4B}{}^{AC} \cdot \bb^{+B} \cdot \bb_C + \cdots,
\nonumber \\
&& \brs \bb_A = \dtwo \bb_A 
- \frac{1}{3!} gf_{3A}{}^{BCD} \cdot \bh_B \cdot \bh_C \cdot \bh_D
- gf_{4A}{}^{BC} \cdot \bh_B \cdot \bb_C + \cdots,
\nonumber \\
&& \brs \bb^{+A} = \dtwo \bb^{+A}
- gf_2^A
- gf_{4B}{}^{CA} \cdot \bb^{+B} \cdot \bh_C + \cdots,
\label{nlbrstrans}
\end{eqnarray}
where the ellipses represent the terms stemming from higher $S_1^{(k)}$
($k \geq 2$).

For all the antifields vanishing, we finally arrive at 
the action (\ref{nlaction}) with gauge symmetry (\ref{gaugebrs})
in the previous section.
Accordingly, if we set
\begin{eqnarray}
&& g f_1^{AB} = W^{AB},
\qquad 
g f_2^{A} = V^{A},
\nonumber \\
&& g f_{3A}{}^{BCD} = - X_{A}{}^{BCD},
\qquad 
g f_{4A}{}^{BC} = - U_{A}{}^{BC},
\end{eqnarray}
Eq.(\ref{fidentity}) coincides with Eq.(\ref{defu}).
%

\section{Dimensional Reduction of 3D Theories}
\noindent
In this section, we dimensionally reduce the 3D nonlinear gauge theory
based on Courant algebroid to 2D theory.

Let $X$ be a three-dimensional manifold
with a coordinate $(\tau, \sigma, \rho)$
and $M$ be a target manifold
of a smooth map $\phi:X \to M$  with local
coordinate expression $\{\phi^i\}$.
We also have a vector bundle $E$ over $M$
with $A^a$ a section of $T^*X \otimes \phi^*(E^*)$.

We can construct 3D topological gauge field theory of the Schwarz type 
in terms of $\phi^i$ and $A^a$
\cite{Izaw,Ikeda:2000yq,Ikeda:2001fq,Ikeda:2002wh,Ikeda:2002qx}.
For that purpose,
we further introduce
$B_{1a}$ as a section of $T^*X \otimes \phi^*(E)$
and $B_{2i}$ as a section of $ \wedge^2 T^*X \otimes \phi^*(T M)$.
Hereafter, the letters $a, b, \cdots$ represent
indices on the fiber of $E$
and $i, j, \cdots$ represent indices on $M$ and $TM$.

\subsection{3D theory based on Lie algebroid}
\noindent
As a simplest example, let us first try
3D nonlinear BF theory with nonlinear gauge symmetry
based on Lie algebroid
or in the case with the target $M$
a Poisson manifold equipped with a Poisson
bivector $\omega^{ij}(\phi)=-\omega^{ji}(\phi)$ and $E=TM$.

A Lie algebroid over a manifold $M$ is a vector bundle 
$E \rightarrow M$ with a Lie algebra structure on the 
space of the sections $\Gamma(E)$ defined by the 
bracket $[e_1, e_2]$ for $e_1, e_2 \in \Gamma(E)$
and a bundle map (the anchor)
$\rho: E \rightarrow TM$ satisfying the following properties:
\begin{eqnarray}
&& {\rm{for \ any}} \quad e_1, e_2 \in \Gamma(E), \quad
[\rho(e_1), \rho(e_2)] = \rho([e_1, e_2]);
\nonumber \\
&& {\rm{for \ any}} \quad e_1, e_2 \in \Gamma(E), 
\ F \in C^{\infty}(M), 
\qquad [e_1, F e_2] = F [e_1, e_2] + (\rho(e_1) F) e_2.
  \label{liealgdef}
\end{eqnarray}

If $E=TM$ and $M$ is a Poisson manifold, 
a Poisson bivector $\omega(\phi)$ defines a Lie algebroid: 
We take $\{ e^i \}$ as a local basis of $\Gamma(TM)$ and
let a local expression of a Poisson bivector 
be $\omega^{ij}(\phi)=-\omega^{ji}(\phi)$. 
Then we can define a Lie algebroid by the following equations:
\begin{eqnarray}
&& [e^i, e^j] = \frac{\partial \omega^{ij}(\phi)}
{\partial \phi^k} e^k 
\nonumber \\ 
&& \rho(e^i) = \omega^{ij}(\phi) \frac{\partial }{\partial \phi^j}.
\end{eqnarray}

We now adopt 3D nonlinear gauge theory
with an action
\cite{Izaw,Ikeda:2000yq}
\begin{eqnarray}
&& S = S_0 +S_1;
\nonumber \\
&& 
S_0 =  \int_{X} \left( B_{1i} \wedge d A{}^i - B_{2i} \wedge d
\phi^i \right),
\nonumber \\
&& S_1 
= \int_{X} \left( \omega{}^{ij} (\phi) B_{2i} B_{1j} 
+  \frac{1}{2} \omega_i^{jk}(\phi)
 A{}^i B_{1j} B_{1k} \right),
\label{bfla}
\end{eqnarray}
where we have defined
\begin{equation}
 \omega_i^{jk}(\phi)
 \equiv \frac{\partial \omega^{jk}(\phi)}{\partial \phi^i}.
\end{equation}

We consider dimensional reduction of the theory
from the three-dimensional manifold $X=\Sigma \times S^1$ to 
the two-dimensional manifold $\Sigma$.
Namely, all the fields are set independent of the coordinate $\rho$
of $S^1$ with $\int_{S^1}d\rho=1$:
\begin{eqnarray}
&& \phi^i = \phizero^i,  \nonumber \\
&& A{}^i = \aone^i + \azero^i d \rho,  \nonumber \\
&& B_{1i} = \bone_i + \bezero_i d \rho,  \nonumber \\
&& B_{2i} = \btwo_{i} + \beone_{i} d \rho, 
\end{eqnarray}
where
$\phizero^i$ is a reduction of $\phi^i$,
$\aone^i$ and $\bone_i$ are $1$-forms, 
$\azero^i$ and $\bezero_i$ are $0$-forms,
$\btwo_{i}$ is a $2$-form, and $\beone_{i}$ is a $1$-form
in two dimensions. 

Then the action (\ref{bfla}) is reduced to the following action: 
\begin{eqnarray}
S 
&=&  \int_{X} \left( \bone_i \wedge \dtwo \azero^i
+ \aone^i \wedge \dtwo \bezero_i 
+ \beone_i \wedge \dtwo \phizero^i \right) d \rho
+ d \left( \aone^i \bezero_i d \rho \right)
\nonumber \\
&& + \biggl(
  \omega{}^{ij}(\phizero) ( \btwo_i \bezero_j - \beone_i \bone_j )
+  \frac{1}{2} \omega_i^{jk}(\phizero) ( 2 \aone^i \bone_j \bezero_k 
+ \azero^i \bone_j \bone_k )
\biggr) d \rho,
\label{bflareduct}
\end{eqnarray}
which can be also obtained through the action (\ref{nlaction})
by letting
\begin{eqnarray}
&& h_A = ( \beone_i, \aone^j, \bone_{k}), \nonumber \\
&& \Phi^A = ( \phizero^i, \bezero_j, \azero^k), \nonumber \\
&& \btwo_A = ( \btwo_i, 0, 0); \nonumber \\
&& W^{AB} = \left( \begin{array}{ccc}
0 & 0 & -\omega{}^{in} \\
0 & 0 & \omega_{j}^{np} \bezero_p  \\
\omega{}^{lk} & - \omega_{k}^{mp} \bezero_p & 
\omega_{p}^{kn} \azero^p \\
\end{array}
\right), 
\nonumber \\
&& V^A = (\omega{}^{ip} \bezero_p, 0, 0),
\end{eqnarray}
for
${}^{\displaystyle A}
=({}^{\displaystyle i},\, {}_{\displaystyle j}\,,\, {}^{\displaystyle k})$.

The 3D nonlinear gauge theory based on Lie algebroid 
from $X$ to $M$ reduces to a 
2D nonlinear gauge theory with two-forms 
from $\Sigma$ to $TM \oplus T^*M$
as a sigma model by dimensional reduction.

When the manifold $M$ is symplectic or the $\omega^{ij}$
is invertible,
we may eliminate the fields $\bezero_j$ and $\btwo_i$
by means of the equations of motion.
Then we further obtain a Poisson sigma model as the reduced theory
(that is, the 2D theory above is reducible) with
\begin{eqnarray}
&& h_A = ( \beone_i, \bone_{k}), \nonumber \\
&& \Phi^A = ( \phizero^i, \azero^k), \nonumber \\
&& W^{AB} = \left( \begin{array}{cc}
0 & -\omega{}^{in} \\
\omega{}^{lk} & \omega_{p}^{kn} \azero^p \\
\end{array}
\right), 
\end{eqnarray}
for $A =(i, k)$.


\subsection{Nonlinear Chern-Simons theory}
\noindent
We can generally construct nonlinear Chern-Simons theory
with nonlinear gauge symmetry
as a deformation of the Chern-Simons gauge theory.
This nonlinear gauge theory in three dimensions 
has the following action\cite{Ikeda:2002wh}:
\begin{eqnarray}
&& 
S = S_0 +S_1,
\nonumber \\
&& 
S_0 
=  \int_{X} \left( \frac{k_{ab}}{2} A^a \wedge d A^b - B_{2i} \wedge d
\phi^i
\right),
\nonumber \\
&& 
S_1 = \int_{X} \left( f_{1a}{}^i (\phi) A^a B_{2i} 
+  \frac{1}{6} f_{2abc} (\phi) A^a A^b A^c
\right),
\label{csbf}
\end{eqnarray}
where 
$k_{ab}$ is a symmetric constant tensor and
the structure functions $f_1$ and $f_2$ satisfy the identities
\begin{eqnarray}
&& k^{ab} f_{1a}{}^i f_{1b}{}^j = 0, 
\nonumber \\ 
&& \frac{\partial f_{1b}{}^i  }{\partial \phi^j} f_{1c}{}^j 
- \frac{\partial f_{1c}{}^i  }{\partial \phi^j} f_{1b}{}^j 
+ k^{ef} f_{1e}{}^i f_{2fbc} = 0, 
\nonumber \\
&& \left( f_{1d}{}^j \frac{\partial f_{2abc} }
{\partial \phi^j}
- f_{1c}{}^j \frac{\partial f_{2dab}  }{\partial \phi^j}
+ f_{1b}{}^j \frac{\partial f_{2cda}  }{\partial \phi^j}
- f_{1a}{}^j \frac{\partial f_{2bcd}  }{\partial \phi^j} 
\right) \nonumber \\
&& 
+ k^{ef} (f_{2eab} f_{2cdf} 
+ f_{2eac} f_{2dbf}
+ f_{2ead} f_{2bcf} )
= 0.
\label{identity3}
\end{eqnarray}
We assume that $k_{ab}$ is nondegenerate or invertible.

A Courant algebroid \cite{Courant}
is a vector bundle $E \rightarrow M$
with a nondegenerate symmetric bilinear form
$\bracket{\cdot}{\cdot}$ 
on the bundle, a bilinear operation $\circ$ on $\Gamma(E)$ (the
space of sections on $E$), and a bundle map (called the anchor) 
$\rho: E \rightarrow TM$ satisfying the following properties:
\begin{eqnarray}
&& 1) \quad e_1 \circ (e_2 \circ e_3) = (e_1 \circ e_2) \circ e_3 
+ e_2 \circ (e_1 \circ e_3), \nonumber \\
&& 2) \quad \rho(e_1 \circ e_2) = [\rho(e_1), \rho(e_2)], \nonumber \\
&& 3) \quad e_1 \circ F e_2 = F (e_1 \circ e_2)
+ (\rho(e_1)F)e_2, \nonumber \\
&& 4) \quad e_1 \circ e_2 = \frac{1}{2} {\cal D} \bracket{e_1}{e_2},
\nonumber \\ 
&& 5) \quad \rho(e_1) \bracket{e_2}{e_3}
= \bracket{e_1 \circ e_2}{e_3} + \bracket{e_2}{e_1 \circ e_3},
  \label{courantdef}
\end{eqnarray}
where 
$e_1, e_2$, and $e_3$ are sections of $E$;
$F$ is a function on $M$;
${\cal D}$ is a map from functions on $M$ to $\Gamma(E)$ 
and is defined by
$\bracket{{\cal D}F}{e} = \rho(e) F$.

If we take a local basis, 
Eq.(\ref{identity3}) is equivalent to the relations 1) to 5) of 
structure functions of a Courant algebroid on a vector bundle
$E$ over $M$:
We take a basis $e^a$ of $\Gamma(E)$.
Then symmetric bilinear form is defined by 
$\bracket{e^a}{e^b} = k^{ab}$.
The bilinear operation is defined by 
$e^a \circ e^b = - k^{ac} k^{bd} f_{2cde} (\phi) e^e$ 
and the anchor is defined by $\rho (e^a)
= - f_{1c}{}^i (\phi) k^{ac} \frac{\partial}{\partial \phi^i}$.

We again consider dimensional reduction of the theory
from the three-dimensional manifold $X=\Sigma \times S^1$ to 
the two-dimensional manifold $\Sigma$:
\begin{eqnarray}
&& \phi^i = \phizero^i,  \nonumber \\
&& A^a = \aone^a + \azero^a d \rho,  \nonumber \\
&& B_{2i} = \btwo_{i} + \beone_{i} d \rho, 
\end{eqnarray}
where
$\phizero^i$ is a reduction of $\phi^i$,
$\aone^a$ is a $1$-form, $\azero^a$ is a $0$-form,
$\btwo_{i}$ is a $2$-form, and $\beone_{i}$ is a $1$-form
in two dimensions. 

Then the action (\ref{csbf}) is reduced to the following action:
\begin{eqnarray}
S
&=&  \int_{X} \left( k_{ab} \aone^a \wedge \dtwo \azero^b 
+ \beone_i \wedge \dtwo \phizero^i \right) d \rho
+ d \left( \frac{k_{ab}}{2} \aone^a \azero^b d \rho
\right)
\nonumber \\
&& + \left( f_{1a}{}^i (\phizero) ( \aone^a \beone_i +
\azero^a \btwo_i )
+  \frac{1}{2} f_{2abc} (\phizero) \aone^a \aone^b \azero^c 
\right) d \rho,
\label{acreduct}
\end{eqnarray}
which can be also obtained through the action (\ref{nlaction})
by letting
\begin{eqnarray}
&& h_A = ( \beone_i, k_{ab} \aone^b), \nonumber \\
&& \Phi^A = ( \phizero^i, \azero^a), \nonumber \\
&& \btwo_A = ( \btwo_i, 0); \nonumber \\
&& W^{AB} = \left( \begin{array}{cc}
0 & -k^{bc} f_{1c}{}^i \\
 k^{ac} f_{1c}{}^j & k^{ad}k^{be} f_{2dec} \azero^c 
\\
\end{array}
\right),
\nonumber \\
&& V^A = (f_{1a}{}^i \azero^a, 0),
\end{eqnarray}
for $A = (i, a)$.
The corresponding gauge symmetry is given by 
\begin{eqnarray}
&& U_j{}^{Ai} = 
\left( 
0, k^{ab} \frac{\partial f_{1b}{}^i}{\partial \phizero^j}
\right)
\nonumber \\
&& U_C{}^{AB} = 0, 
\qquad \mbox{otherwise};
\nonumber \\
&& X_j{}^{abc} = - k^{ad} k^{be} k^{cf}
\frac{\partial f_{2def}}{\partial \phizero^j},
\nonumber \\
&& X_D{}^{ABC} = 0, 
\qquad \mbox{otherwise},
\end{eqnarray}
which satisfies the identities (\ref{defu})
due to the identities (\ref{identity3}).

The 3D nonlinear Chern-Simons theory from $X$ to $M$
reduces to a 2D nonlinear gauge theory with two-forms 
from $\Sigma$ to $E$ as a sigma model by dimensional reduction.

\subsection{3D nonlinear BF theory}
\noindent
We can also construct
3D nonlinear BF theory with nonlinear gauge symmetry
as a deformation of the BF gauge theory in three dimensions.
This nonlinear gauge theory in three dimensions 
has the following action
\cite{Ikeda:2002qx}:
\begin{eqnarray}
&& S = S_0 +S_1;
\nonumber \\
&& 
S_0 =  \int_{X} \left( B_{1a} \wedge d A{}^a - B_{2i} \wedge d
\phi^i
\right),
\nonumber \\
&& S_1 
= \int_{X} ( f_{1a}{}^i (\phi) A{}^a B_{2i} 
+ f_{2}{}^{ib} (\phi) B_{2i} B_{1b} 
+  \frac{1}{6} f_{3abc} (\phi) A{}^a A{}^b A^c
\nonumber \\ 
&& \quad \quad \ 
+  \frac{1}{2} f_{4ab}{}^c (\phi) A{}^a A{}^b B_{1c}
+  \frac{1}{2} f_{5a}{}^{bc} (\phi) A{}^a B_{1b} B_{1c}
+  \frac{1}{6} f_6{}^{abc} (\phi) B_{1a} B_{1b} B_{1c}
),
\label{bf}
\end{eqnarray}
where 
the structure functions $f_1, \cdots, f_6$ satisfy the identities
\begin{eqnarray}
&& 
f_{1}{}_e{}^i f_{2}{}^{je} + f_{2}{}^{ie} f_{1}{}_e{}^j = 0, 
\nonumber \\
&& 
- \frac{\partial f_{1}{}_c{}^i}{\partial \phi^j} f_{1}{}_b{}^j
+ \frac{\partial f_{1}{}_b{}^i}{\partial \phi^j} f_1{}_c {}^j
+ f_1{}_e{}^i f_{4bc}{}^e + f_{2}{}^{ie} f_{3ebc} = 0, 
\nonumber \\
&&
 f_1{}_b{}^j \frac{\partial f_{2}{}^{ic}}{\partial \phi^j} 
- f_{2}{}^{jc} \frac{\partial f_1{}_b{}^i}{\partial \phi^j} 
+ f_1{}_e{}^i  f_{5b}{}^{ec} - f_{2}{}^{ie} f_{4eb}{}^c = 0, 
\nonumber \\
&&
- f_{2}{}^{jb} \frac{\partial f_{2}{}^{ic}}{\partial \phi^j} 
+ f_{2}{}^{jc} \frac{\partial f_{2}{}^{ib}}{\partial \phi^j} 
+ f_1{}_e{}^i f_{6}^{ebc} + f_{2}{}^{ie} f_{5e}{}^{bc} = 0, 
\nonumber \\
&&
- f_1{}_{[a}{}^j \frac{\partial f_{4bc]}{}^d}{\partial \phi^j} 
+ f_{2}{}^{jd} \frac{\partial f_{3abc}}{\partial \phi^j} 
+ f_{4e[a}{}^d f_{4bc]}{}^{e} + f_{3e[ab} f_{5c]}{}^{de} = 0, 
\nonumber \\
&&
- f_1{}_{[a}{}^j \frac{\partial f_{5b]}{}^{cd}}{\partial \phi^j} 
- f_{2}{}^{j[c} \frac{\partial f_{4ab}{}^{d]}}{\partial \phi^j} 
+ f_{3eab} f_6{}^{ecd} 
+ f_{4e[a}{}^{[d} f_{5b]}{}^{c]e} + f_{4ab}{}^e f_{5e}{}^{cd} = 0, 
\nonumber \\
&&
- f_1{}_a{}^j \frac{\partial f_{6}{}^{bcd}}{\partial \phi^j} 
+ f_{2}{}^{j[b} \frac{\partial f_{5a}{}^{cd]}}{\partial \phi^j} 
+ f_{4ea}{}^{[b} f_6{}^{cd]e} + f_{5e}{}^{[bc} f_{5a}{}^{d]e} = 0, 
\nonumber \\
&&
- f_{2}{}^{j[a} \frac{\partial f_{6}{}^{bcd]}}{\partial \phi^j} 
+ f_6{}^{e[ab} f_{5e}{}^{cd]} = 0, 
\nonumber \\
&&
- f_1{}_{[a}{}^j \frac{\partial f_{3bcd]}}{\partial \phi^j} 
+ f_{4[ab}{}^{e} f_{3cd]e} = 0.
\label{3dJacobi2}
\end{eqnarray}
Note that the 3D theory in subsection 4.1 is an example of this action.

As is the case in the previous subsection,
if we take a local basis, 
Eq.(\ref{3dJacobi2}) is equivalent to the relations of 
structure functions of a Courant algebroid on a vector bundle 
$E \oplus E^*$ over $M$:
Symmetric bilinear form $\bracket{\cdot}{\cdot}$ 
is defined from the natural pairing of $E$ and $E^*$.
That is,
$\bracket{e_a}{e_b} = \bracket{e^a}{e^b} = 0$ and 
$\bracket{e_a}{e^b} = \delta_a{}^b$
if $\{e_a\}$ is a basis of sections of $E$
and $\{e^a\}$ is that of $E^*$.
The bilinear form $\circ$ and the anchor $\rho$ are represented 
as follows:
\begin{eqnarray}
&& e{}^a \circ e{}^b = - f_{5c}{}^{ab}(\phi) e^c
- f_{6}{}^{abc}(\phi) e_{c}, \nonumber \\
&& e^a \circ e_{b} = - f_{4bc}{}^{a}(\phi) e^c
+ f_{5b}{}^{ac}(\phi) e_{c}, \nonumber \\
&& e_{a} \circ e_{b} = - f_{3abc}(\phi) e^c
- f_{4ab}{}^{c}(\phi) e_{c}, \nonumber \\
&& \rho(e^a) = - f_{2}{}^{ia}(\phi) 
\frac{\partial}{\partial \phi^i}, \nonumber \\
&& \rho(e_a) = - f_{1a}{}^i(\phi) 
\frac{\partial}{\partial \phi^i}.
\end{eqnarray}

We again consider dimensional reduction of the theory
from the three-dimensional manifold $X=\Sigma \times S^1$ to 
the two-dimensional manifold $\Sigma$:
\begin{eqnarray}
&& \phi^i = \phizero^i,  \nonumber \\
&& A{}^a = \aone^a + \azero^a d \rho,  \nonumber \\
&& B_{1a} = \bone_a + \bezero_a d \rho,  \nonumber \\
&& B_{2i} = \btwo_{i} + \beone_{i} d \rho, 
\end{eqnarray}
where
$\phizero^i$ is a reduction of $\phi^i$,
$\aone^a$ and $\bone_a$ are $1$-forms, 
$\azero^a$ and $\bezero_a$ are $0$-forms,
$\btwo_{i}$ is a $2$-form, and $\beone_{i}$ is a $1$-form
in two dimensions. 

Then the action (\ref{bf}) is reduced to the following action: 
\begin{eqnarray}
S 
&=&  \int_{X} \left( \bone_a \wedge \dtwo \azero^a
+ \aone^a \wedge \dtwo \bezero_a 
+ \beone_i \wedge \dtwo \phizero^i \right) d \rho
+ d \left( \aone^a \bezero_a d \rho \right)
\nonumber \\
&& + \biggl( f_{1a}{}^i ( \aone^a \beone_i + \azero^a \btwo_i) 
+ f_2{}^{ib} ( \btwo_i \bezero_b - \beone_i \bone_b )
+  \frac{1}{2} f_{3abc} \aone^a \aone^b \azero^c 
\nonumber \\
&& 
+  \frac{1}{2} f_{4ab}{}^c ( \aone^a \aone^b \bezero_c 
- 2 \aone^a \bone_c \azero^b )
+  \frac{1}{2} f_{5a}{}^{bc} ( 2 \aone^a \bone_b \bezero_c 
+ \azero^a \bone_b \bone_c )
\nonumber \\
&& 
+  \frac{1}{2} f_6^{abc} \bone_a \bone_b \bezero_c
\biggr) d \rho
\nonumber \\
&=&  \int_{\Sigma} \bone_a \wedge \dtwo \azero^a
+ \aone^a \wedge \dtwo \bezero_a 
+ \beone_i \wedge \dtwo \phizero^i
+ f_{1a}{}^i \aone^a \beone_i 
+ f_2{}^{ib} \bone_b \beone_i 
\nonumber \\
&& 
+  \frac{1}{2} (f_{3abc} \azero^c + f_{4ab}{}^c \bezero_c )
\aone^a \aone^b 
+  ( - f_{4ab}{}^c \azero^b  + f_{5a}{}^{cb} \bezero_b ) \aone^a \bone_c 
\nonumber \\
&& 
+  \frac{1}{2} ( f_{5a}{}^{bc} \azero^a + f_6^{abc} \bezero_a)
\bone_b \bone_c
+ (f_{1b}{}^i \azero^b + f_2{}^{ia} \bezero_a ) \btwo_i,
\label{bfreduct}
\end{eqnarray}
which can be also obtained through the action (\ref{nlaction})
by letting
\begin{eqnarray}
&& h_A = ( \beone_i, \aone^a, \bone_{b}), \nonumber \\
&& \Phi^A = ( \phizero^i, \bezero_a, \azero^b), \nonumber \\
&& \btwo_A = ( \btwo_i, 0, 0); \nonumber \\
&& W^{AB} = \left( \begin{array}{ccc}
0 & - f_{1c}{}^i &  -f_{2}{}^{id} \\
f_{1a}{}^j & f_{3ace} \azero^e + f_{4ac}{}^e \bezero_e  & 
- f_{4ae}{}^d \azero^e + f_{5a}{}^{de} \bezero_e  \\
f_{2}{}^{jb} & f_{4be}{}^c \azero^e - f_{5b}{}^{ce} \bezero_e & 
f_{5e}{}^{bd} \azero^e + f_{6}^{bde} \bezero_e  \\
\end{array}
\right), 
\nonumber \\
&& V^A = (f_{1b}{}^i  \azero^b + f_2{}^{ia} \bezero_a, 0, 0),
\end{eqnarray}
for
${}^{\displaystyle A} 
=({}^{\displaystyle i},\, {}_{\displaystyle a}\,,\, {}^{\displaystyle b})$.
The corresponding gauge symmetry is given by
\begin{eqnarray}
&& U_j{}^{Ai} = \left( 
0, \frac{\partial f_{1a}{}^i}{\partial \phizero^j},
\frac{\partial f_{2}{}^{ib}}{\partial \phizero^j} \right)
\nonumber \\
&& U_C{}^{AB} = 0, 
\qquad \mbox{otherwise};
\nonumber \\
&& X_{jabc} = - \frac{\partial f_{3abc}}{\partial \phizero^j},
\nonumber \\
&& X_{jab}{}^c = - \frac{\partial f_{4ab}{}^c}{\partial \phizero^j},
\nonumber \\
&& X_{ja}{}^{bc} = - \frac{\partial f_{5a}{}^{bc}}{\partial \phizero^j},
\nonumber \\
&& X_j{}^{abc} = - \frac{\partial f_6{}^{abc}}{\partial \phizero^j},
\nonumber \\
&& X_D{}^{ABC} = 0, 
\qquad \mbox{otherwise},
\end{eqnarray}
where complete antisymmetrization for $X_j{}^{ABC}$
with respect to the indices $ABC$ should be understood.
This again satisfies the identities (\ref{defu})
due to the identities (\ref{3dJacobi2}).

The 3D nonlinear BF theory from $X$ to $M$
reduces to a 
2D nonlinear gauge theory with two-forms 
from $\Sigma$ to $E \oplus E^*$ as a sigma model 
by dimensional reduction.

\section{Conclusion}
\noindent
We have investigated the 2D nonlinear gauge theory with two-forms,
which is obtained as the consistent deformation
of 2D topological BF gauge theory.

Dimensional reduction of 3D nonlinear gauge theory
based on Courant algebroid
such as nonlinear Chern-Simons theory and 3D nonlinear BF theory
yields such 2D nonlinear gauge theory with two-forms.
If it is reducible as is considered at the end of section 2,
we obtain the Poisson sigma model based on Lie algebroid
with a reduced target space.
Namely, the reduction of the base space is accompanied by that of
the target space structure with the Courant algebroid
reduced to the Lie algebroid.%
\footnote{In specific cases, higher-dimensional theory itself
can be directly based on the Lie algebroid
\cite{Izaw,Ikeda:2000yq}
prior to dimensional reduction (see the first
case study in the previous section).}

We have analyzed the algebroid defined by Eq.(\ref{defu}) in 
terms of the Batalin-Vilkovisky algebra in section 3.
Further analyses of algebraic and geometric structures of this type of
theories would shed more light on 
relations between topological field theories and algebroids, 
including a Lie algebroid and a Courant algebroid.

The web of topological gauge field theories of the Schwarz type
may be connected by deformations and reductions,
as is exemplified in the cases of two- and three-dimensional
nonlinear gauge theories in this paper.
It can be also deformed to nontopological theories
\cite{Izawa}
and constitutes an intriguing arena in the space of field theories
from a deformation theory perspective.

\section*{Acknowledgements}

We would like to thank T.~Strobl for valuable comments.


\section*{Appendix}
\noindent
The Batalin-Vilkovisky antibracket
\cite{Bat}
for functions $F(\varphi, \varphi^+)$ and 
$G(\varphi, \varphi^+)$ of the fields
and antifields on the base space $X$ is defined by
\begin{eqnarray}
\abv{F}{G} \equiv \frac{F \rd}{\partial \varphi} 
\frac{\ld G}{\partial \varphi^+}
- (-1)^{(n+1)\deg \varphi}
\frac{F \rd}{\partial \varphi^+} \frac{\ld G}{\partial \varphi},
\label{anti}
\end{eqnarray}
where $n=\dim X$ and ${\rd}/{\partial \varphi}$ and ${\ld}/{\partial
\varphi}$ are the right and
left differentiations 
with respect to $\varphi$, respectively, which satisfy
\begin{eqnarray}
\frac{\ld F}{\partial \varphi} =  (-1)^{(\gh F + \gh \varphi) \gh \varphi 
+ (\deg F + \deg \varphi) \deg \varphi}
\frac{F \rd}{\partial \varphi}.
\label{lrdif}
\end{eqnarray}
${\rd}/{\partial \varphi^+}$ and 
${\ld}/{\partial \varphi^+}$ have similar definitions.
When $F$ and $G$ are functionals of the fields $\varphi$
and antifields $\varphi^+$,
the antibracket is given by
\begin{eqnarray}
\abv{F}{G} \equiv \int_{X}
\left(
\frac{F \rd}{\partial \varphi} \frac{\ld G}{\partial \varphi^+}
- (-1)^{(n+1)\deg \varphi} 
\frac{F \rd}{\partial \varphi^+} \frac{\ld G}{\partial \varphi}
\right).
\label{antif}
\end{eqnarray}

The antibracket satisfies the following identities:
\begin{eqnarray}
&& \abv{F}{G} 
= -(-1)^{(\deg F + n)(\deg G + n) + (\gh F + 1)(\gh G + 1)} \abv{G}{F},
\nonumber \\
&& \abv{F}{GH} = \abv{F}{G} H + (-1)^{(\deg F + n)\deg G + (\gh F +1) \gh G}
G \abv{F}{H},
\nonumber \\
&& \abv{FG}{H} = F \abv{G}{H} 
+ (-1)^{\deg G(\deg H + n) + \gh G(\gh H + 1) } \abv{F}{H} G,
\label{antibra}
\\
&& (-1)^{(\deg F + n)(\deg H + n) + (\gh F + 1)(\gh H + 1) } 
\abv{F}{\abv{G}{H}}
+ {\rm cyclic \ permutations} = 0.
\nonumber
\end{eqnarray}
We also note that
\begin{eqnarray}
&& FG = (-1)^{\gh F \gh G + \deg F \deg G} G F, \nonumber \\
&& d(FG) = dF G + (-1)^{\deg F} F dG.
\label{FGpro}
\end{eqnarray}

In order to simplify cumbersome sign factors, we introduce
{super product},
{super antibracket}, and {super differentiation}
\cite{Cat}.%
\footnote{{The super product} is called {the dot product} in
Ref.\cite{Ikeda:2001fq,Ikeda:2002wh,Ikeda:2002qx,Cat}.}

Let us define the {\it super product} by
\begin{eqnarray}
F \cdot G \equiv  (-1)^{\gh F \deg G} FG.
\label{dotpro}
\end{eqnarray}
We obtain the following identities from Eq.(\ref{FGpro}):
\begin{eqnarray}
&& F \cdot G = (-1)^{|F||G|} G \cdot F, \nonumber \\
&& d (F \cdot G) = d F \cdot G + (-1)^{|F|} F \cdot d G,
\end{eqnarray}
where $|F| \equiv \gh F + \deg F$ denotes the total degree of $F$.

The {\it super antibracket} is defined by
\begin{eqnarray}
\sbv{F}{G} \equiv (-1)^{(\gh F + 1) (\deg G + n)}
(-1)^{\gh \varphi (\deg \varphi + n) + n}
\abv{F}{G}.
\label{dotantibran}
\end{eqnarray}
Then the following identities are obtained from Eq.(\ref{antibra}):
\begin{eqnarray}
&& \sbv{F}{G} = -(-1)^{(|F| + n + 1)(|G| + n + 1)} \sbv{G}{F},
\nonumber \\
&& \sbv{F}{G \cdot H} = \sbv{F}{G} \cdot H
+ (-1)^{(|F| + n + 1)|G|} G \cdot \sbv{F}{H},
\nonumber \\
&& \sbv{F \cdot G}{H} = F \cdot \sbv{G}{H}
+ (-1)^{|G|(|H| + n + 1)} \sbv{F}{H} \cdot G,
\nonumber \\
&& (-1)^{(|F| + n + 1)(|H| + n + 1)} \sbv{F}{\sbv{G}{H}}
+ {\rm cyclic \ permutations} = 0.
\end{eqnarray}
That is, the super antibracket
provides a graded Poisson bracket on {\it superfields}.

We further define the {\it super differentiation} by
\begin{eqnarray}
&& \frac{\ld }{\partial \varphi} \cdot F 
\equiv (-1)^{\gh \varphi \deg F}
\frac{\ld F}{\partial \varphi}, \nonumber \\
&&F \cdot \frac{\rd }{\partial \varphi}
\equiv (-1)^{\gh F \deg \varphi}
\frac{F \rd}{\partial \varphi}.
\end{eqnarray}
We can define the {\it super differentiation} with respect to 
$\varphi^+$ in a similar manner.
Owing to Eq.(\ref{lrdif}), we obtain
\begin{eqnarray}
\frac{\ld }{\partial \varphi} \cdot F =   
(-1)^{(|F| + |\varphi|) |\varphi|}
F \cdot \frac{\rd }{\partial \varphi}.
\label{lrdiff}
\end{eqnarray}

\newpage

\newcommand{\bibit}{\sl}


\vfill\eject
\end{document}